\documentclass[12pt,preprint]{aastex}
\pdfoutput=1

\renewcommand {\deg}   {\mbox{$^\circ$}}

\newcommand   {\kms}   {\mbox{km\,s$^{-1}$}}
\renewcommand {\ga}    {\mbox{\rlap{\hbox{\lower5pt\hbox{$\sim$}}}\hbox{$>$}}}
\renewcommand {\la}    {\mbox{\rlap{\hbox{\lower5pt\hbox{$\sim$}}}\hbox{$<$}}}

\begin{document}
\pagenumbering{arabic} 
\def\kms {\hbox{km{\hskip0.1em}s$^{-1}$}} 
\voffset=-0.8in

\def\msol{\hbox{$\hbox{M}_\odot$}}
\def\lsol{\hbox{$\hbox{L}_\odot$}}
\def\kms{km s$^{-1}$}
\def\Blos{B$_{\rm los}$}
\def\etal   {{\it et al.}}                     
\def\psec           {$.\negthinspace^{s}$}
\def\pasec          {$.\negthinspace^{\prime\prime}$}
\def\pdeg           {$.\kern-.25em ^{^\circ}$}
\def\degree{\ifmmode{^\circ} \else{$^\circ$}\fi}
\def\ut #1 #2 { \, \textrm{#1}^{#2}} 
\def\u #1 { \, \textrm{#1}}          
\def\nH {n_\mathrm{H}}
\def\ddeg   {\hbox{$.\!\!^\circ$}}              
\def\deg    {$^{\circ}$}                        
\def\le     {$\leq$}                            
\def\sec    {$^{\rm s}$}                        
\def\msol   {\hbox{$M_\odot$}}                  
\def\i      {\hbox{\it I}}                      
\def\v      {\hbox{\it V}}                      
\def\dasec  {\hbox{$.\!\!^{\prime\prime}$}}     
\def\asec   {$^{\prime\prime}$}                 
\def\dasec  {\hbox{$.\!\!^{\prime\prime}$}}     
\def\dsec   {\hbox{$.\!\!^{\rm s}$}}            
\def\min    {$^{\rm m}$}                        
\def\hour   {$^{\rm h}$}                        
\def\amin   {$^{\prime}$}                       
\def\lsol{\, \hbox{$\hbox{L}_\odot$}}
\def\sec    {$^{\rm s}$}                        
\def\etal   {{\it et al.}}                     
\def\la{\lower.4ex\hbox{$\;\buildrel <\over{\scriptstyle\sim}\;$}}
\def\ga{\lower.4ex\hbox{$\;\buildrel >\over{\scriptstyle\sim}\;$}}
\def\refitem{\par\noindent\hangindent\parindent}
\oddsidemargin = 0pt \topmargin = 0pt \hoffset = 0mm \voffset = -17mm
\textwidth = 160mm  \textheight = 244mm
\parindent 0pt
\parskip 5pt

\shorttitle{Star formation near Sgr A*}
\shortauthors{}

\title{ALMA Detection of  Bipolar Outflows: Evidence\\
for Low Mass Star Formation within 1pc of  Sgr A*} 
\author{F. Yusef-Zadeh$^1$, M. Wardle$^2$, D. Kunneriath$^3$, 
M. Royster$^1$, A. Wootten$^3$ \&  D. A. Roberts$^4$}

\affil{$^1$CIERA, Department of Physics and Astronomy Northwestern University, Evanston, IL 60208}
\affil{$^2$Dept of Physics and Astronomy,  Research Centre for Astronomy, Astrophysics\\
and Astrophotonics, Macquarie University, Sydney NSW 2109, Australia}
\affil{$^3$National Radio Astronomy Observatory,  Charlottesville, VA 22903}
\affil{$^4$Fort Worth Museum of Science and History, Fort Worth, TX 76107}


\begin{abstract} 
We report the discovery of 11 bipolar outflows within a projected distance of 1pc from Sgr A* based 
on deep ALMA observations of $^{13}$CO, H30$\alpha$ and SiO (5-4) lines with sub-arcsecond and 
$\sim1.3$ \kms\, resolutions.  These unambiguous signatures of young protostars manifest as 
approaching and receding lobes of dense gas swept up by the jets created during the formation and 
early evolution of stars. 
The lobe 
masses and momentum transfer rates are consistent with young protostellar outflows found throughout 
the disk of the Galaxy. 
The mean dynamical age of the outflow population is  estimated to be
$6.5^{+8.1}_{-3.6}\times10^3$ years. 
The  rate of star formation is 
$\sim5\times10^{-4}$\msol\,yr$^{-1}$ assuming a mean stellar mass of $\sim0.3$ \msol. 
This discovery provides evidence that star formation is taking place within 
clouds surprisingly close to Sgr A*, perhaps due to events that compress the host cloud,  creating 
condensations with sufficient self-gravity to resist tidal disruption by Sgr A*.  
Low-mass star formation over the past few billion years at this level 
would contribute significantly to the stellar mass 
budget in the central few pc of the Galaxy. The presence of many dense clumps of molecular material 
within 1pc of Sgr A* suggests 
that star formation could take place in the immediate vicinity of supermassive black holes in 
the nuclei of external galaxies. 
\end{abstract}

\keywords{accretion, accretion disks --- black hole physics --- Galaxy: center}

\section{Introduction}

The $4\times10^6\msol$ black hole at the center of the Milky Way, Sgr A*, is expected to suppress star formation in 
nearby interstellar clouds because of tidal disruption by the black hole's gravitational field 
\citep{1993ApJ...408..496M}. Nevertheless, objects resembling dust-enshrouded young stars 
\citep{2004ApJ...602..760E,2013ApJ...778...92Y,2015ApJ...801L..26Y} and photo-evaporative flows from their disks 
\citep{2015ApJ...801L..26Y,2016ApJ...819...60Y} have been identified within a few light years of Sgr A*. Clear 
identification of the nature of these objects has been hampered by the Galactic center's distance, 30 magnitudes of 
foreground extinction, and stellar crowding.

The density required for self-gravity to overcome the tidal field of Sgr A*, $\mathrm{\sim 10^7 (r/pc)^{-3}}$ 
cm$^{-3}$, far exceeds the density in molecular clouds ($\mathrm{\la10^6}$\,cm$^{-3}$) and so is believed to 
inhibit star formation via gravitational collapse of dense regions of clouds.  By contrast, the concentration 
of $\sim$100 OB stars
within 
$\sim$0.4 pc ($\sim10''$) of Sgr~A* \citep{2004ApJ...602..760E,2006ApJ...643.1011P,2009ApJ...690.1463L} is  a few million years old, and is thought to have formed in 
its present disk-like geometry.  Approximately seven to eight million years ago, a migrating 
molecular cloud swept past Sgr A* and captured material forming a gaseous disk orbiting Sgr A* \citep{2007MNRAS.379...21N,2008ApJ...683L..37W,2012ApJ...750L..38W,2008Sci...321.1060B,2012ApJ...749..168M}.  The disk 
became gravitationally unstable as it settled and cooled,  and fragmented into stars.

A variety of observations suggest that star formation is 
happening inside clouds within  a parsec of Sgr A* despite inhibition  by tidal disruption. 
These include compact sources suggestive of embedded young 
stars \citep{2004ApJ...602..760E,2013A&A...551A..18E,2013ApJ...778...92Y,2015ApJ...809...10Y} and 
their precursor  dense 
cloud cores \citep{2016PASJ...68L...7T}.  Radio and millimeter continuum sources 
are also found 
that resemble the photo-evaporative flows 
from protoplanetary disks
\citep{2015ApJ...801L..26Y,2016ApJ...819...60Y,2017MNRAS.467..922Y} seen in the Orion and 
Trifid nebulae \citep{1998Natur.396..343R}, along with water and methanol 
masers \citep{2008ApJ...683L..37W,2015ApJ...808...97Y}, and SiO (5-4) emission with 
luminosities and line widths matching those seen in jets from young stars in molecular clouds 
in the Galactic disk \citep{2013ApJ...767L..32Y,2015ApJ...801L..26Y}. 
However, none of these are regarded as conclusive, because conditions 
at the Galactic center make it difficult to distinguish dust-enshrouded high-mass YSOs from dusty evolved stars or 
low-mass YSOs from starless dusty cores \citep{2010ApJ...721..395F,2012Natur.481...51G,2013A&A...551A..18E}.  Furthermore, the physical and chemical conditions that create  masers and  SiO emission could 
plausibly arise in colliding flows in the complex of gas orbiting Sgr A* rather than in shock waves driven by outflows from YSOs.

Here, we report the discovery of a population of bipolar outflows, an unambiguous signature of young protostars within  0.8 parsec (pc) of  Sgr A*.

\section{Data Reduction and Results}

Deep observations were carried out with 
Atacama Large Millimeter/submillimeter Array (ALMA) 
with high spectral resolution, spatial 
resolution and sensitivity to image $^{13}$CO (J=2--1) and SiO (5--4) molecular lines and, in one 
instance, the H30$\alpha$ radio recombination line (RRL) of hydrogen in the inner pc of the Galaxy.  
To image $^{13}$CO (2-1), H30$\alpha$ RRL and SiO (5-4) emission at 220.398, 231.901, 217.105 GHz 
respectively, we used calibrated 
Band 6 archival 
data with a single $\sim26''$ field of view in Cycle 3 (project code 2015.A.00021.S) observed on July 
12/13, 2016.

We imaged the ALMA continuum data after self-calibration, using CASA
version 4.7.2.  After all spectral windows were combined, phase
self-calibration solutions were derived and subsequently applied for
three solution intervals: scan length, 30.25 seconds, and the
integration time.  In addition, a phase and amplitude self-calibration
solution for an interval equal to the integration time was applied.
After continuum subtraction, these phase and amplitude solutions were
transferred to the $^{13}$CO and H30$\alpha$ line spectral window with
spectral resolutions of 1.33 and 1.26 \kms, respectively. For the final
line imaging, we combined data taken in both days and used Brigg's
weighting with robust parameters of 0.5 and 2, giving spatial
resolutions of $0.49''\times0.39''$ (PA=-68.9$^\circ$) and
$0.51''\times0.39''$ (PA=-79.9$^\circ$), respectively.  Briggs parameters
of 2 and 0.5 are used to weigh the  {\it uv} data \citep{1999ASPC..180..127B}.
Briggs value of 2 (0.5) is 
closer to natural  (uniform) weighting and  more sensitive to  extended (compact) 
features. 
We imaged the
spectral line data cubes with channel widths of 1.33 and 1 \kms,
respectively.
We also used archival and published HCN (1-0)
\citep{2005ApJ...622..346C} data with spatial (spectral) resolutions of
$5.1\times2.7''$ (7.2 \kms),  CS (5-4) and SiO (6-5) data
with $0.76''\times0.59''$ (18.75 \kms) resolution.


We identified eleven sources with physical characteristics (size, mass, age, energetics) 
of young star formation activity 
within 0.8 pc of the black hole. The integrated flux density, angular size and position angle (PA) 
of individual sources are listed 
in Table 1a and our estimates of physical parameters are given in Table 1b.  
One object, Bipolar 1 (BP1),  with compact bipolar CO-emission lies 
at a projected distance of 
$\sim0.6$ pc (15.2$''$) SW of Sgr A*, as shown 
in Figure 1a. 
The  NW and SE lobes are blue- and red-shifted by $\sim$2 \kms\, with respect to each other, 
connected by  
intermediate velocity gas at  +113 \kms (green). The axis of this feature has a 
PA of $-42^\circ$.1, and resembles a jet connecting the two lobes. 
The centroids of the lobes of BP1 are separated by 
$\sim1.45''$, equivalent to $\sim1.2\times10^4$ AU at the 8 kpc distance of the Galactic center.   
The $^{13}$CO lobes 
likely arise from the material swept up by a jet driven outflow from a central young stellar object. The total mass 
swept up in the NW and SE lobes is estimated to be $\sim$0.19 and 0.12 \msol, respectively. 
The equivalent momentum 
deposition rates are $\sim$ 1180 and 820 $\lsol$, respectively, consistent with an outflow 
arising  from a low mass  protostar.  


BP1 should therefore be embedded in a molecular cloud.  
A CS (5-4) line intensity map is presented in Figure 1b, 
taken  from an ALMA observation \citep{2017A&A...603A..68M},  
which identifies  
a  dense and filamentary molecular cloud with narrow linewidths  
within 0.5$''$ of BP1 at a velocity of 103 \kms\, 
with  a linewidth of 18.75 \kms.  
CS (5-4) has a high  critical density of 
5.4$\times10^6$ cm$^{-3}$ suggesting a dense cloud, most likely 
the parent cloud from which the CO lobes  have been swept by BP1's protostellar jet 
(see below other tracers of the parent cloud). 
Figure 1c shows contours of $^{13}$CO emission between +110 and +118 \kms. The blue lobe has a
dominant central maximum and is elongated along the symmetry axis, as expected of a bipolar
molecular outflow. 

Additional  evidence of   a molecular cloud near BP1 can be found in Figures 1e,f,g 
showing  a grayscale 
$^{13}$CO line intensity  between 106 and 114  \kms, 
a color image  between +110 and +117 \kms\, 
and  grayscale 
contours of HCN (1-0) 
emission between +110 and +118 \kms, respectively. 
The systemic velocity of BP1 is similar 
to those shown in these figures suggesting that the parent cloud is associated 
with the inner edge of the ring orbiting Sgr~A*. 

The interferometric HCN, CO  and CS data  suffer from the lack of short
UV spacings, thus diffuse, extended and low density molecular emission is  filtered out. 
In addition, the inner region of the molecular ring is traced 
by ionized gas associated with the mini-spiral. 
Absorption of molecular gas against the associated radio continuum 
is significant in the inner region (Marr et al. 1992). 
In spite of these limitations, there is clear evidence 
that BP1 is
embedded within the  inner edge of the
molecular ring and is generated by a jet pushing on the surrounding molecular material.  
BP1 lies at the edge of a long filamentary structure (see Fig. 1e)
at the interface of the 
molecular ring and the Western Arc, a layer of ionized gas and dusty material 
at a velocity of between +70 and +90 \kms.

We also detect a  
faint unresolved 
compact continuum source (BP1-226GHz) at 226 GHz between the two lobes at 0.74 $\pm 0.19$ mJy with
a resolution of 0.48$''\times0.38''$ at 
$\alpha, \delta (J2000)=17^h\, 45^m\, 38^s.923, -29^\circ\, 00'\, 30''.604$. 
BP1-226GHz  is likely to be thermal dust emission associated 
with the disk of the protostar driving the outflow (see Fig. 1h). The disk mass, 
assuming  a dust temperature of 50 K,  and dust opacity $\kappa_\nu = 0.068 $\,cm$^2$\,g$^{-1}$ is $\sim$0.05 
\msol. 
In addition, we find a 3.6$\mu$m source very close to the position of the 226 GHz source, as shown 
by a circle in Figure 1h. The relative displacement of the IR source may 
simply reflect the astrometric errors in aligning ALMA and VLT images as these were 
taken six years apart (S. Gillessen, private communication). 
We tentatively detect a  possible counterpart to BP1  at longer 
IR wavelengths (N band).  
Future study of this source including its proper motion is needed to confirm the 
association of the mid-IR source with the 226 GHz  continuum source. 

Evolved stars also drive outflows, but they are generally isotropic because of the relatively low 
specific angular momentum of stellar rotation. In protoplanetary nebulae and related 
systems, a central binary provides orbital angular momentum and powers collimated outflows. However, 
these tend to have rather more extreme velocities \citep{2017ApJ...835L..13S} and are traced by 
ionized gas.

Another bipolar source (BP2) is shown 
in Figure 2a with  
 contours of blue and red-shifted $^{13}$CO emission from two lobes and a
third source to the north. These sources are shown superimposed on a grayscale 34 GHz continuum
image \citep{2015ApJ...809...10Y}. 
The region where the bipolar source BP2 lies is surrounded  by ionized features with similar radial velocities based 
on radio recombination line observations (see Fig. 2a). 
Figure 2b shows the distribution of CS (5-4) emission adjacent to the source. The
presence of very dense gas at a similar velocity to  the lobes indicate that BP2 is associated with a dense  cloud
traced by CS (5-4) emission. Figure 2c shows the CO emission arising from velocities between 91-102 \kms\, in 16
adjacent channel maps. There are three velocity components located to the west of the N arm of the ionized mini-spiral
and the parent CS cloud.  
There is some mixing of blue and red-shifted velocity components as shown in the southern
lobes. The peak emission from blue and red-shifted lobes show a velocity difference of 4 \kms\, with respect to each
other centered at 97 \kms\, and the lobe's centroids are spatially separated by$\sim1''.29$.  The position-velocity
diagram of the lobes along a cut that passes through the peaks of blue and red-shifted components is shown in Figure
2d. The continuous velocity gradient, $\sim$3 \kms\, arcsecond$^{-1}$ suggests that the blue and red-shifted
components are associated. The velocity and spatial distributions suggests that the overlapping velocity components
represent a bipolar outflow with the red-shifted component dominating the emission. A faint velocity component,
1.5$''$NE of the red-shifted component, has a peak velocity of 95 \kms\, and appears as a distinct source.  It is
possible that this velocity component is another bipolar outflow source that is spatially and kinematically 
unresolved.



The remaining nine BP sources,  BP3 to BP11,  are displayed  in Figure 3a-i, respectively. 
Briefly, all sources  show two lobes of emission 
with receding and approaching velocities, some coincide with a peak in CS (5-4) emission, giving a 
molecular core mass of $\sim$1 \msol (BP2, BP6, BP9) if we assume that the gas density is the 
critical density of CS (5-4) line emission, $2\times10^7$ cm$^{-3}$, 
some others, e.g., BP3, show multiple transitions of SiO 
emission with a jet-like morphology. The SiO (8-7) emission from BP3 requires a critical density of 
$2\times10^7$ cm$^{-3}$ (BP3) suggesting a highly compressed jet-like molecular material by 
protostellar outflows. Some sources coincide with a 226 GHz continuum source, tracing a protostellar 
disk (BP8) or with infrared excess sources with SEDs that are consistent with those of young stellar 
objects (BP3, BP4) and some show irradiated lobes of ionized gas by the external radiation field of 
the Galactic center (BP7, BP8).

Figure 4a shows the positions and elongated axis position angles of all eleven sources superimposed on 
an image of $^{13}$CO emission.  The red and blue  colors represent the red- and blue-shifted CO 
emission, respectively, excluding velocities between -50 and 50 \kms. It is clear that the molecular 
gas in the ring and its interior provides a reservoir of molecular gas to explain the origin of the 
bipolar outflow sources. CO emission from the interior of the molecular gas is consistent with recent 
measurements \citep{2009ApJ...695.1477M,2017A&A...603A..68M}, suggesting dense molecular gas 
self-shields itself against strong ionizing radiation from massive stars. Evidence for dense 
molecular gas inside the molecular ring is also viewed in the color image of Figure 4b, which shows 
CO emission between -250 and +250 \kms. It is clear that $^{13}$CO emission fills the interior of the 
molecular ring in contrast to the long held view that the central cavity inside the molecular ring is 
devoid of any molecular gas \citep{2005ApJ...622..346C,2005ApJ...620..287H}.

\section{Discussion}

Overall, the population of outflow sources presented here constitutes overwhelming 
evidence that star formation is taking place in clouds near the black hole despite the disruptive 
effect of its strong tidal field.  Squeezing of clouds by 
the  black hole's strong  tides  can contribute to the  collapse of clouds
\citep{2014MNRAS.444.1205J} as well as 
 high interstellar pressure in the inner few pc of the galaxy, 
$\sim 2\times10^9$\,K\,cm$^{-3}$ stabilizes the progenitor clouds against disruption.  However, they 
are rendered gravitationally unstable when subject to pressures 1-2 orders of magnitude higher 
associated with a triggering event such as collision with the gas of the minispiral, a high ram 
pressure of an approaching cloud toward Sgr A* or a jet emerging from Sgr A* 
\citep{2017IAUS..324..111Y}; Wardle and Yusef-Zadeh 2017, in preparation). 
Both the ionized and molecular gas in the ring  and the mini-spiral 
provide strong  external pressure against molecular cores 
and  trigger star formation.


The mass of the individual outflow lobes is estimated by assuming that $^{13}$CO emission is 
optically thin and is in LTE with $T\sim 100-300$\,K.  
The estimated dynamical ages and 
momentum transfer rates depend on the unknown inclinations of the outflows with respect to the line 
of sight.  We account for this by assuming that the intrinsic distribution of outflow axis 
orientation in three dimensions is isotropic.  This allows us to estimate the mean dynamical age of 
the population as $6.5^{+8.1}_{-3.6}\times10^3$ years (95\% confidence interval), somewhat shorter than 
the orbital period around Sgr A*, 17000 (r/0.5pc)$^{1.5}$ years where r is the distance from Sgr A*.  
The estimated masses and momentum transfer rates are consistent with young protostellar 
outflows found throughout the disk of the Galaxy (Dunham et al. 2014).


The distribution of the outflow axis orientations from all 11 sources is shown as line segments on Figure 4a. The 
observed and derived properties of individual sources are listed in Table 1a,b.  We note that the outflow sources are 
generally surrounded by ionized and molecular gas and located within $\sim$1 pc of Sgr~A*. Individual maps of CS 
(5-4) emission shows that seven of the 11 outflow sources are associated with molecular clouds. In spite of the low 
spectral resolution of CS data, the molecular core coincides with the outflow source and there is nearby dense gas 
that is swept by the jet from the protostar in almost all of the observed sources described here. In addition, out of 
11 sources, the outflow axes of six sources are within a cone that passes through Sgr A* with PAs between 26$^\circ$ 
and 63$^\circ$.  It is suggestive that there is anisotropy in the distribution of outflows sources. However, this 
anisotropy is not significant statistically.  Kuipers's test applied to the distribution of PAs shows no statistical 
significance because of the small sample size. Confirmation of this correlation has to await for a larger sample of 
sources. If the distribution were anisotropic, large scale magnetic field in the interior of the molecular ring and 
the mini-spiral or a large-scale collimated outflow from Sgr A* compressing gas clouds could be responsible. We note 
a number of cometary radio and infrared sources, elongated nonthermal radio continuum features at a 
PA$\sim50-60^\circ$ \citep{2007A&A...469..993M,2010A&A...521A..13M,2012ApJ...758L..11Y,2016ApJ...819...60Y}.  These 
features are interpreted as interaction sites of a collimated mildly relativistic jet from Sgr A* with the atmosphere 
of stars and the nonthermal Sgr A East shell.
An independent set of polarization measurements at near-IR\citep{2015A&A...576A..20S}
shows that the position angle of Sgr A* varies
over a range within the cone where most bipolar sources are detected. This distribution
is consistent with a collimated outflow producing
head-tail sources and possibly compressing gas, thus inducing star formation along the direction where most
bipolar outflows have been detected.

Notably, detection of bipolar flows is the only means to identify low-mass YSOs, which are otherwise 
too faint to be detected given the 30 magnitudes of foreground visual extinction and 8\,kpc distance 
to the Galactic center and in any case difficult to identify in the IR given the crowded field toward 
the Galactic center. Follow up observations will census the population allowing studies of the 
initial mass function (IMF) and a comparison with star formations in less extreme environments.  The 
11 detected low-mass YSOs formed within the last 6500 years imply a star formation rate 
$\sim5\times10^{-4}$\msol\,yr$^{-1}$ assuming a mean stellar mass of $\sim0.3$ \msol. 
The rate would be higher if the mean mass is larger.
If this is 
reflective of the mean rate of low-mass star formation over the past few billion years then this mode 
of star formation contributes significantly to the stellar mass budget in the central few pc of the 
Galaxy.

Our detection of bipolar sources in this apparently hostile environment suggests that star formation 
activity might also be higher than expected in the central molecular zone within the inner few 
hundred pc of the Galaxy; further searches should reveal the extent and distribution of star 
formation there. 
The measurement presented here shows that the central cavity of the molecular ring 
is not only filled with ionized gas but also with dense molecular gas, providing fuel for star 
formation \citep{2013ApJ...767L..32Y,2016PASJ...68L...7T,2017A&A...603A..68M}. 
Future  ALMA observations  provide prospects for 
characterizing  the gas properties inside the molecular ring. 
A census of low-mass 
star formation at the Galactic center will provide a more accurate estimate of star formation rate 
and the IMF near supermassive black holes.
Follow up observations will 
characterize the population of outflows, allowing studies of the initial mass function (IMF) and a 
comparison with star formation in less extreme environments. 
This perspective is critical for 
understanding star formation occurring in the center of external galaxies that host supermassive 
black holes. 


{\bf{Acknowledgements}}
This work is partially supported by the grant AST-0807400 from the NSF. 
The National Radio Astronomy Observatory is a facility of the National Science Foundation operated under cooperative 
agreement by Associated Universities, Inc.
This paper makes use of the following ALMA data: ADS/JAO.ALMA\#2011.0.00005.SV. ALMA is a partnership of ESO 
(representing its member states), NSF (USA) and NINS (Japan), together with NRC (Canada), NSC and ASIAA (Taiwan), and 
KASI (Republic of Korea), in cooperation with the Republic of Chile. The Joint ALMA Observatory is operated by ESO, 
AUI/NRAO and NAOJ.
We thank R. Arendt  and the referees for making useful comments.  

\bibliographystyle{apj}

\begin{table}
\centering
\caption{
{\it {a Top}}
Entries
give the source name, the coordinates, the flux density, the extent of the blue- and red-shifted  lobes 
in terms of the beam solid angle, the velocity difference of the lobes, the angular separation of the lobes 
in arcseconds, the center velocity  and the position angle of the source in degrees.
{\it {b Bottom}}
Physical parameters of molecular  outflows with 
entries giving  the source name, the mass and hydrogen number density of individual lobes, 
the total mass of the lobes, the jet length and the outflow force.
}
\medskip
\begin{tabular}{ccccccccccc}
\hline
       &     \multicolumn{2}{c}{J2000 coord}        & \multicolumn{2}{c}{\underline{Red Lobe}} &
\multicolumn{2}{c}{\underline{Blue Lobe}} &   &      & &  \\
       & RA(s) & Dec($''$) & F$\Delta$v & $\Omega$                    &
F$\Delta$v                & $\Omega$      & $\Delta$v & $\Delta\theta$  & v$_{\mathrm{center}}$ & PA\\
{\scriptsize Source} & {\scriptsize $17^h45^m$}       & {\scriptsize $-29^\circ00^\prime$} & {\scriptsize mJy km 
s$^{-1}$} & {\scriptsize beam}            &
{\scriptsize mJy km s$^{-1}$}         & {\scriptsize beam}        & {\scriptsize {km s$^{-1}$}} 
& $''$ & {\scriptsize km s$^{-1}$} & {\scriptsize deg.} \\
\hline
BP1 	& 38.917 	& 30.72 	& 215   	& 2.9 	& 135 	& 2.4 	& 2 	& 1.45 
 	& 113.0 	& -42.1	\\
BP2 	& 40.417 	& 15.48 	& 1140  	& 3.6 	& 450   	& 1.9 	& 4 	& 1.29 	& 97.0 	& 26.6	\\
BP3 	& 40.495 	& 29.11 	& 7.8   	& 1.5 	& 10.3  	& 1.1 	& 1 	& 0.74    	& 163.5 	& 51.3  	\\
BP4 	& 40.840 	& 21.03	& 54  	& 1.5 	& --  		& -- 		& 1 	& 0.17  	& 197.5 	
& 54.3  	\\
BP5 	& 39.869 	& 22.24 	& 63  	& 1.4  	& 116   	& 1.6  	& 4 	& 0.41  	& -93.0 
	& -88.9	\\
BP6 	& 39.343	& 44.27	& 2489 	& 6.5 	& 1198 	& 5.1 	& 5 	& 1.01  	& 68.5 	& 5.1	\\
BP7 	& 39.440	& 32.82	& 22.2 	& 2.9 	& -- 		& -- 		& 1 	& 1.80  	& -199.5 	& 56.1	\\
BP8$_{HII}$		& 39.638	& 31.28	& 6.2 	& 1.2 	& 5.1 	& 1.1 	& 13 & 0.81  	& -81.8 	& -50.7 	\\
BP9 	& 40.414	& 30.41	& 155 	& 2.9 	& 510 	& 2.1 	& 2 	& 3.28  	& -59 	& -50.3	\\
BP10	& 39.296	& 29.39	& 171 	& 1.2 	& 170 	& 1.2 	& 3 	& 0.44  	& -57 	& 62.8	\\
BP11	& 40.553	& 16.62	& 42.7 	& 1.5 	& 44.7 	& 1.35 	& 1 	& 0.39  	& 163.5 	& 22.5	\\
\hline
\end{tabular}
\medskip
\begin{tabular}{ccccccccc}
\hline
       	& \multicolumn{2}{c}{\underline{Red Lobe}} & \multicolumn{2}{c}{\underline{Blue Lobe}} & 
\underline{Total} & & &\\
       & Mass  & n$_{H}$                           & Mass & n$_{H}$                           & Mass              
 & jet length & Force\\
       &       &           &       &           &       &   $\times (\sin\,i)$ &  $\times(\cos^2i/\sin\,i)$ \\
Source & \msol & cm$^{-3}$ & \msol & cm$^{-3}$ & \msol & $10^{17}$ cm & \lsol\, /\,c \\
\hline
   BP1 &     0.19 &  2.6e+05 &     0.12 &  2.1e+05 &     0.32      & 0.87  & 5.7e+02  \\
   BP2 &        1.0 &  9.8e+05 &     0.41 &    1.0e+06 &      1.4  & 0.77  &  1.1e+04  \\
   BP3 &    0.007 &  2.5e+04 &   0.0093 &  5.3e+04 &    0.016      &0.44   &  1.4e+01  \\
   BP4 &    0.049 &  1.7e+05 &       -- &       -- &       --      &0.10   &       --  \\
   BP5 &    0.057 &  2.2e+05 &      0.10 &  3.4e+05 &     0.16     &0.25   &  4.1e+03  \\
   BP6 &      2.2 &  8.9e+05 &      1.1 &  6.1e+05 &      3.3      &0.60   &  5.4e+04  \\
   BP7 &     0.020 &  2.6e+04 &       -- &       -- &       --     &1.1    &       --  \\ 
   BP8$_{HII}$ &   0.0056 &  2.8e+04 &   0.0046 &  2.6e+04 &  0.01 &0.48   &  1.4e+04  \\
   BP9 &     0.14 &  1.8e+05 &     0.46 &  9.9e+05 &      0.60     &2.0    &  4.8e+02  \\
  BP10 &     0.15 &  7.7e+05 &     0.15 &  7.6e+05 &     0.31      &0.26   &  4.1e+03  \\
  BP11 &    0.039 &  1.4e+05 &     0.04 &  1.7e+05 &    0.079      &0.23   &  1.3e+02  \\
\hline
\end{tabular}
\end{table}


\begin{figure}
\center
\caption{(see Fig. 1 next page)
The relative coordinates (0,0) in Figures 1-4 coinicide with the position of 
Sgr A* at $\alpha, \delta (J2000)=17^h\, 45^m\, 40^s.038, -29^\circ\, 00'\, 28''.069$. 
{\it (a)}
The blue and red-shifted lobes are detected to the NE and  SW 
and  the green color
represents the intermediate velocity range.
The intensity range is between -1 and 10 mJy beam$^{-1}$. The cross
coincides with the position of a  compact 226 GHz source (see (h)).
{\it (b)}
A color image of CS(5-4) emission at a velocity of 103 \kms, channel width of 18.75 \kms\, and 
rms noise per channel 0.9 mJy beam$^{-1}$ 
The cross is the same as in (a).
{\it (c)}
Contours of intensity  from 9 channels at 
-1, 1, 2, 3, 4, 5, 6,  8, 10, 12, 14 mJy beam$^{-1}$ \kms\,
with velocities ranging between 110 and 118 \kms. 
A primary beam correction has not been applied. 
The mixing of 
blue and red-shifted CO emission from the two lobes is most likely due to low 1.33 \kms\, spectral resolution.
{\it (d)}
PV diagram constructed along the axis of the symmetry of BP1
at PA$\sim-42^\circ.6$  connecting the two lobes. The color bar to the right
shows the intensity of the $^{13}$CO emission corrected for the primary beam.
{\it (e)} 
$^{13}$CO line intensity integrated over 106 and 114 \kms. 
{\it (f)} 
A true color  $^{13}$CO image at velocities 
between 110-112, 113-114 and 115-117  \kms, in red, green and blue, respectively.
The intensity range is between -1 and 5 mJy beam$^{-1}$.
{\it (g)}
Contours of HCN (1-0) line intensity integrated over 
105 and 118 \kms\,
with levels (-1, -0.5, 0.5, 1,.., 3, 4,...10)$\times0.67$ Jy \kms. 
This data set  is taken from 
interferometric observations (Owens Valley Radio Observatory)
with a resolution of $5.1''\times2.7''$ \citep{2005ApJ...622..346C}. 
{\it (h)}
Contours of 226 GHz emission with a resolution of $0.42''\times0.33''$ (PA=$-74^\circ$) with levels 
-0.1, 0.1, 0.2, 0.3, 0.4, 0.5, 0.6 mJy are superimposed on a grayscale $^{13}$CO image at a velocity of 
113 \kms. A 226 GHz source is detected at the center of BP1. 
{\it (i)}
A composite image of 3.8 $\mu$m (red)  and  $^{13}$CO at 113 \kms\, (green). 
The black circle shows the IR  source which could be 
the counterpart to the 226GHz continuum source seen in (h).    The beam sizes of the CO and CS data are displayed  
on the bottom left corners of (a) and (b), respectively.
}
\end{figure}
\clearpage

\begin{figure}[h]
\includegraphics[scale=0.38,angle=0]{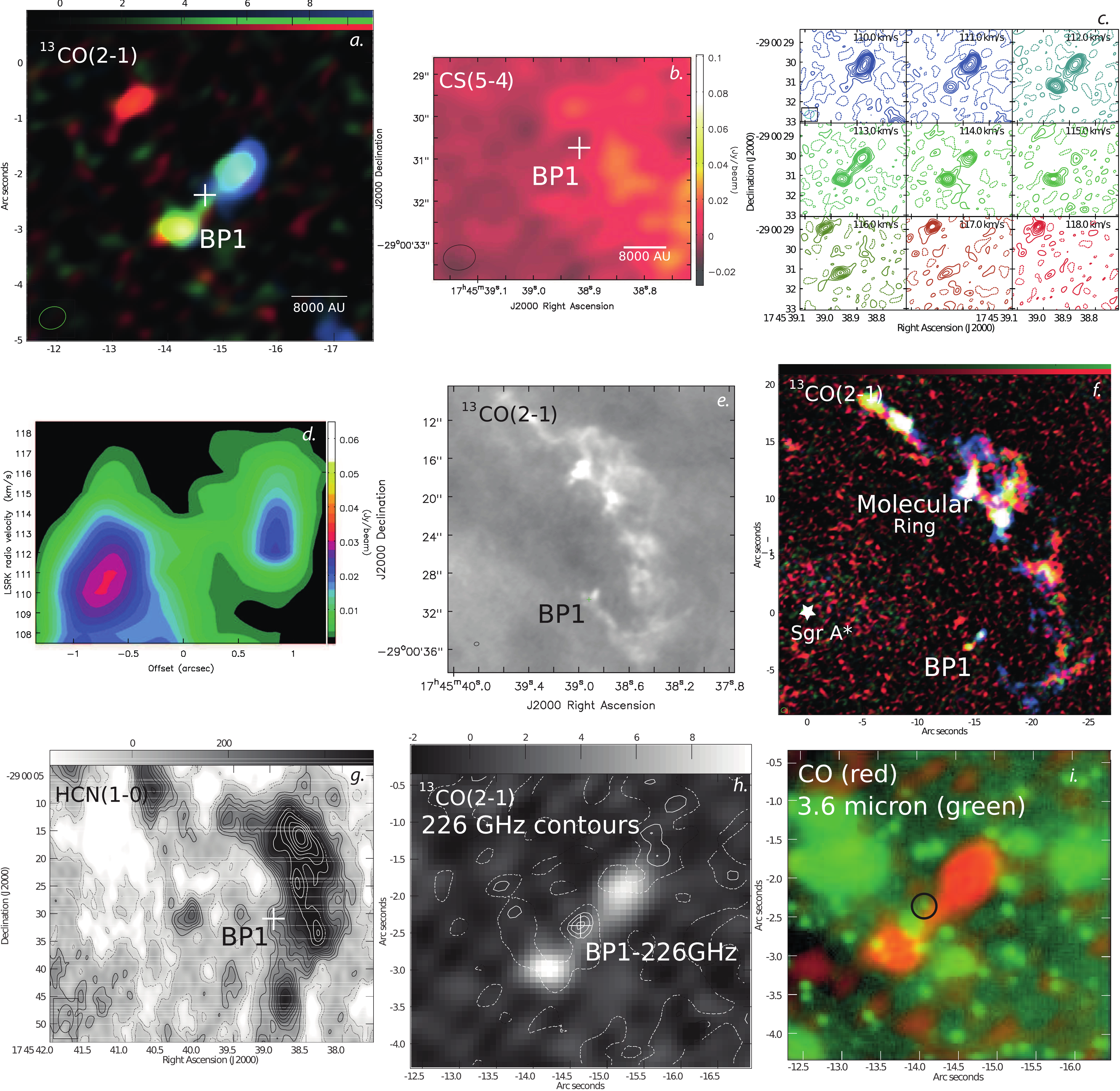}
\end{figure}

\clearpage

\begin{figure}
\caption{(see Fig. 2 next page)
{\it (a)}
Contours of integrated $^{13}$CO emission between 92-96 and 97-101 \kms, shown in blue and red, respectively,
at 2, 2.5, 3, 3.5, 4,  4.5, 5, 6 mJy beam$^{-1}$ \kms\,
are superimposed on a 34 GHz grayscale image with grayscale range between -115 and 100 $\mu$Jy beam$^{-1}$.
{\it (b)}
CS(5-4) emission from a cloud  centered at  84 \kms\, with a linewidth of 18.75 \kms.
The cross shows the center of CO emission from BP1.   The CS(54) rms noise per channel is 0.9 mJy
beam$^{-1,}$ \citep{2017A&A...603A..68M}.
{\it (c)}
Contours of 16 channel maps with levels set at
(-1, 1, 2, 3, 4, 6, 8, 10, 14, 18, 22, 26)$\times3$ mJy beam$^{-1}$.
{\it (d)}
PV diagram of BP2 made with a slice at a PA$=26^\circ.56$.
The beam sizes are shown in bottom left corners in (a) and (b).
}
\end{figure}
\clearpage

\begin{figure}
\includegraphics[scale=0.55,angle=0]{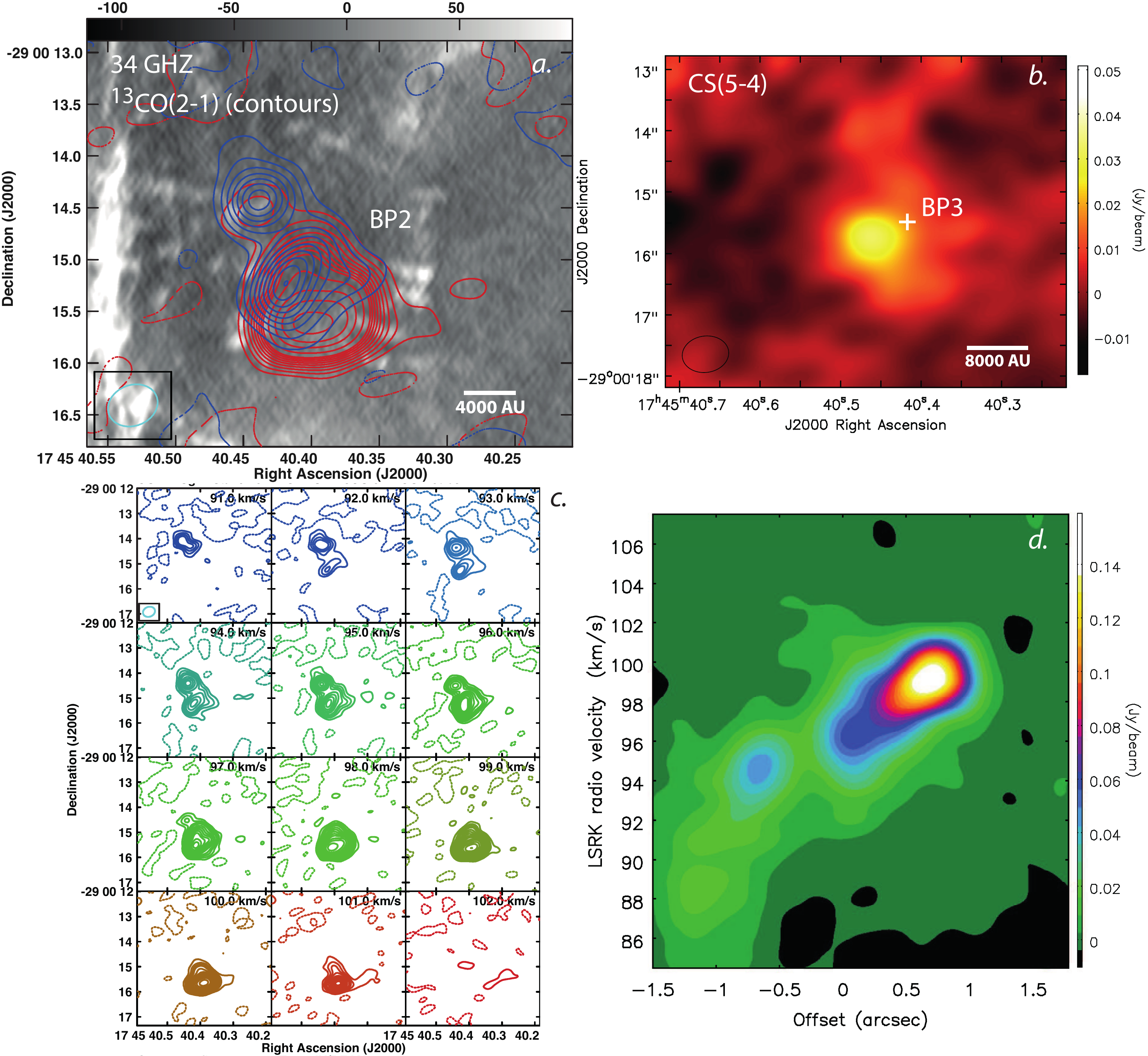}
\end{figure}

\clearpage


\begin{figure}
\caption{(see Fig. 3 next page)
{\it (a)}
A blue- and red-shifted image of BP3 corresponding to
velocities 162-163 and 164-165 \kms\,  respectively. The integrated intensity ranges
between -1 and 2.4 mJy/beam \kms.
{\it (b)}
Contours of red- and blue-shifted CO emission with levels set at -1.5, 1.5, 2, 2.5, 3, 4, 5 and 6 
mJy beam$^{-1}$ \kms at  201 and 196 \kms\, respectively.
{\it (c)}
Contours of  blue and red-shifted $^{13}$CO emission  corresponding to velocities
-96 to -94 and -92 to -90 \kms\, respectively,  with levels at
(-0.4, 0.4, 0.5, 0.6, 0.8, 1, 1.2, 1.4, 1.6, 1.8, 2, 2.2, 2.4, 2.6, 2.8, 3)$\times15$ mJy \kms.
The grayscale image with a range between -100 and 50 $\mu$Jy beam$^{-1}$
shows the head-tail radio source associated with
the supergiant star IRS 7 at the Galactic center.
{\it (d)}
Contours of  blue and red-shifted $^{13}$CO emission  corresponding to velocities
62--69 and 70--76 \kms\, respectively,  with levels at
(-1, 1, 2, 3, 4, 5, 6)$\times100$ mJy \kms.
The grayscale image of the blue-shifted velocities
has a  range between -100 and 100 $\mu$Jy beam$^{-1}$
{\it (e)}
Contours of  integrated $^{13}$CO line  intensity between -200 and -199 \kms\,
at -2, 2,  4, 6, 8  mJy \kms. The color image shows the ionized gas traced by
 H30$\alpha$ RRL (green) and molecular material traced by
  CO emission  (red). The  intensity ranges between -2 and 13 mJy beam$^{-1}$.
{\it (f)}
Contours of  blue (-94.3 to -88 \kms)
and red-shifted (-84.8 to -78.5  \kms)
H30$\alpha$ recombination line  intensity
at velocities between
at (-3, 3, 4, 5, 6, 7, 8, 9, 1)$\times3.2$ mJy \kms with a spatial resolution of
$0.46''\times0.35''$.
The cross coincides with a 226 GHz continuum peak intensity 0.62$\pm0.23$ mJy beam$^{-1}$.
The integrated 226 GHz intensities  from the blue and red-shifted lobes  are 0.63 and 0.54 mJy integrated
over 1.4 and 1.3 times the  solid angle of the synthesized beam which is 0.35$''\times0.22''$.
{\it (g)}
A grayscale image of $^{13}$CO emission at
-56  \kms\, with intensity range between -1 and 20 mJy beam$^{-1}$.
{\it (h)}
A color  image of the blue- and red-shifted lobes of  $^{13}$CO  sources.
Contours of red-shifted component is set at
(-1, 1, 2, 4, 7, 10, 14) $\times6$ mJy  beam$^{-1}$ integrated over velocities between -57 and -54 \kms.
{\it (i)}
Contours of  blue and red-shifted $^{13}$CO intensity
at velocities between 160  to 163 and 164 to 167  \kms\,
at (-3, 3, 4, 5, 6, 7, 8, 9, 10)$\times5$ mJy \kms\,
superimposed on
a 34 GHz continuum image with grayscale range between -100 and 200 $\mu$Jy beam$^{-1}$.
}
\end{figure}

\begin{figure}
\includegraphics[scale=0.4,angle=0]{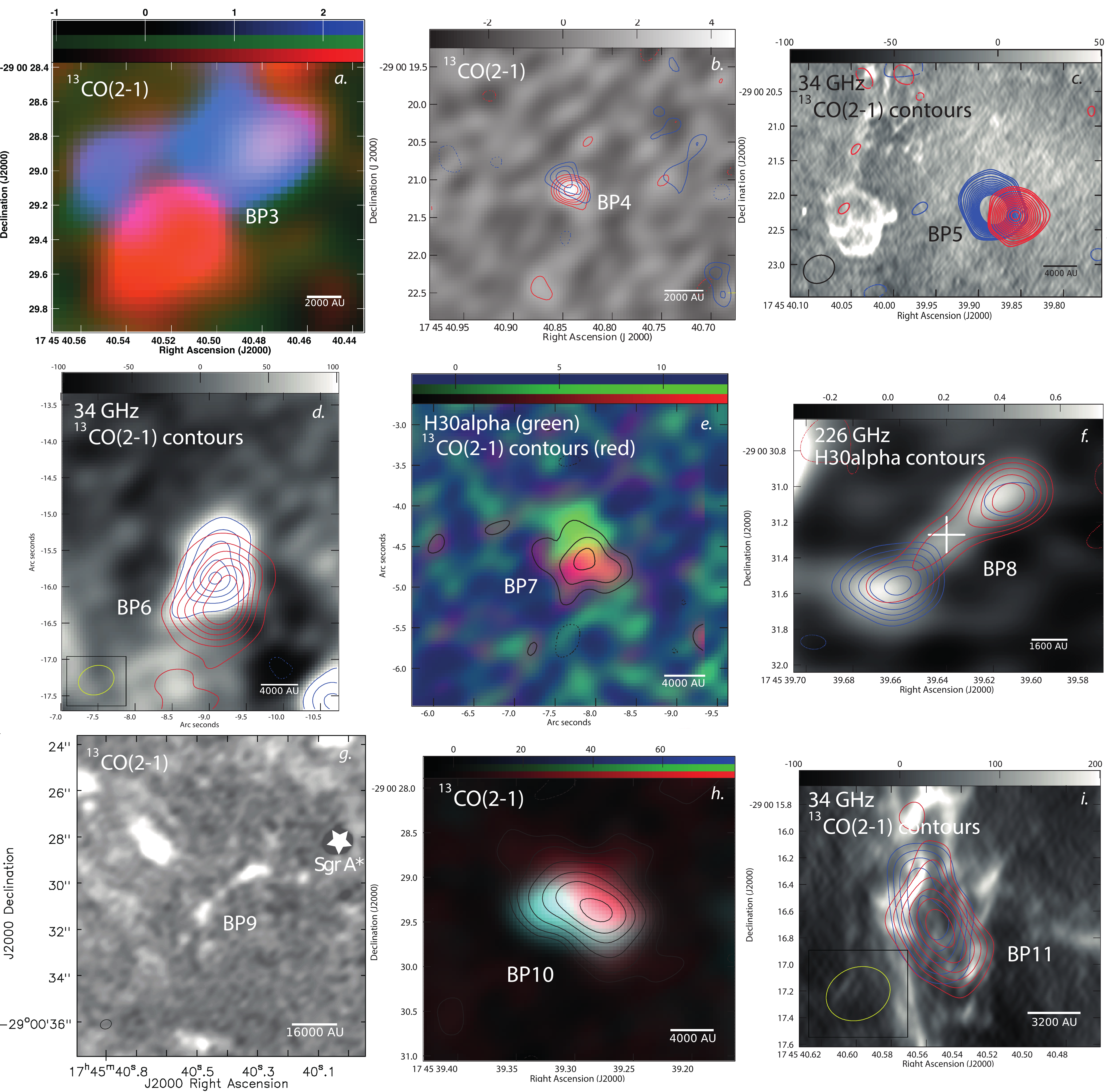}
\end{figure}

\begin{figure}
\caption{(see Fig. 4 next page)
{\it (a)}
The  position and the PA of the symmetry axis  of all eleven    outflow sources are   
drawn as  line segments  and  
are superimposed on  a color image of the distribution of 
 $^{13}$CO emission.  
The $^{13}$CO intensity integrated from  -50 to -200 \kms  and 50 to  200 \kms 
is shown in blue and red, respectively.  The image is not primary beam corrected.  
The flux density range is between -0.66 and 0.66 Jy \kms. The black spot coincides with the position of 
Sgr A*.
The position angle of individual outflow source is tabulated in the last column of Table 1 and is 
given  in degrees east of north. 
{\it (b)}
The distribution of $^{13}$CO showing peak emission at velocities ranging between -250 and 250 \kms.  
The ellipse outlines schematically the molecular ring detected in HCN (1-0) \citep{2005ApJ...622..346C}. 
The images shown in (a) and (b) reveal  that $^{13}$CO  with a wide range of velocities 
 fills the interior of  the molecular ring. 
The CO line intensity ranges  are  -0.01 to 0.1 (red), -0.01 to 0.2 (green) and -0.2 to -0.03 (blue)
 Jy \kms. The image is primary beam corrected.
}
\end{figure}

\clearpage

\begin{figure}
\includegraphics[scale=0.6,angle=0]{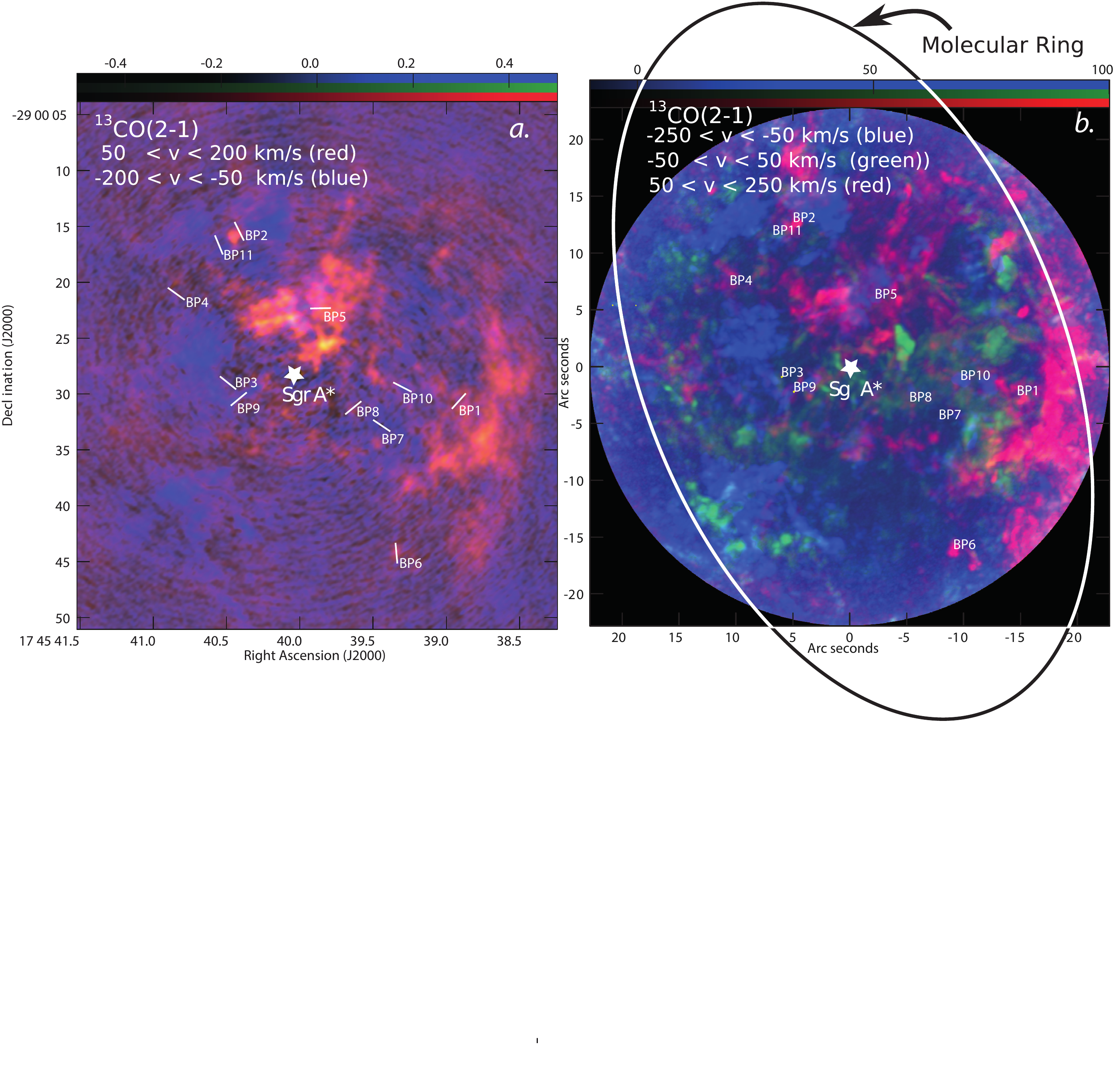}
\end{figure}

\end{document}